\documentclass{appolb}
\usepackage{epsfig}
\usepackage{amssymb,wasysym,cancel,wrapfig}
\begin{document}
\title{Flavour Physics in the Littlest Higgs Model with $T$-Parity:
Effects in the $K$, $B_{d/s}$ and $D$ systems
\thanks{Presented at the Flavianet Workshop on Low energy constraints
on extensions of the Standard Model,
23-27 July 2009,
Kazimierz, Poland
}%
}
\author{Stefan Recksiegel
\address{Technische Universit\"at M\"unchen, Physikdepartment, T31}
}
\maketitle
\begin{abstract}
The Littlest Higgs Model with T parity (LHT) is an interesting 
alternative model for New Physics at the TeV scale. Although Flavour
Physics was not the reason for creating the LHT model, significant
effects (such as large $CP$ violation where not predicted by
the SM) can be created without violating existing experimental bounds.
We study the $B$-, $K$- and especially the $D$-sector.
\end{abstract}


\section{Introduction: Gauge hierarchy in the SM}

\begin{wrapfigure}[3]{o}[2cm]{0pt}
\includegraphics[width=3.5cm,bb=172 345 439 430]{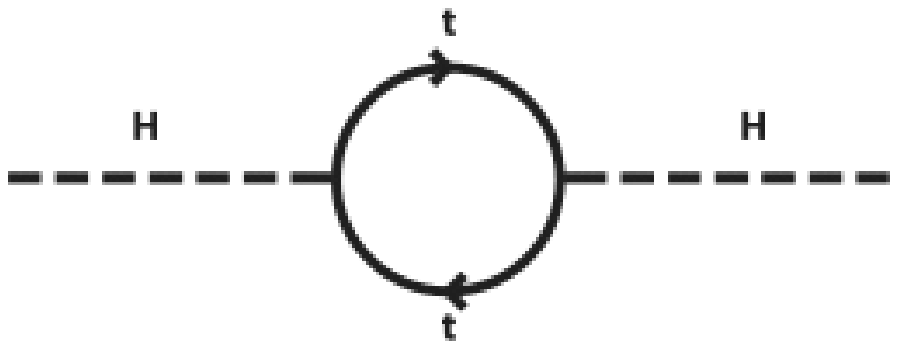}
\end{wrapfigure}
A major problem in the Standard Model (SM) is the { Gauge Hierarchy}
problem, 
Top-loop corrections make the { Higgs mass} unstable,
$ \Delta m_H^2=-{\left|\lambda_t\right|^2\!/ 8\pi^2}
  \left[\Lambda_{UV}^2+\dots\right]$.
To prevent $ m_H \to m_{\rm Planck}$, we need incredible { fine-tuning}.
{ One possible solution is}
{ SUSY}, where the  top-loop is cancelled with a stop-loop,
$ \Delta m_H^2=2{\left|\lambda_s\right|^2\!/ 16\pi^2}
  \left[\Lambda_{UV}^2+\dots\right]$.
It is also possible to lower the Planck mass with { extra dimensions}, 
another possible solution to the Gauge Hierarchy problem is
the Little Higgs mechanism.


\section{The Little(st) Higgs Model (with T parity)}

In the Little Higgs class of models \cite{ArkaniHamed:2001ca},
the { Higgs Boson is a { pseudo-Goldstone} boson of a spontaneously
broken global symmetry. Gauge and Yukawa couplings break the 
symmetry explicitely, but every { single coupling} conserves enough
of the symmetry to keep { the Higgs massless}.}
This way, the radiative corrections to the Higgs mass are
only logarithmically divergent at one loop (and not quadratically
as in the SM).

One popular implementation of the Little Higgs mechanism
is the Littlest Higgs Model \cite{ArkaniHamed:2002qy},
where the { Higgs boson is {a pseudo-Goldstone boson} from
breaking a global $ SU(5)$ symmetry to a global $ SO(5)$
at the scale $ f\sim{\cal O}({\rm TeV})$.}
{ The exact mechanism for symmetry breaking is unspecified,
therefore the Littlest Higgs model is an { effective theory}
valid up to $ \Lambda \sim 4\pi f$.}

There are { 14 Nambu-Goldstone bosons from symmetry breaking:
the { SM Higgs}, new heavy gauge bosons $ W_H^\pm$, $ Z_H$, $ A_H$, 
a scalar triplet $ \Phi$, and a heavy partner for the top quark, $ T$.}
{ In the original Littlest Higgs, custodial $SU(2)$ is
{ broken already at tree level}, then electroweak
precision (EWP) observables demand $ f\gtrsim 2\!-\!3\, {\rm TeV}$,
this leads to rather small 
($10\!-\!20\%$) effects in Flavour Physics.}

By introducing a new  discrete symmetry ({ ``T parity''}),
the Littlest Higgs Model with T parity (LHT) \cite{Cheng:2003ju}
avoids problems with the EWP observables:
Under the new symmetry, all { new} particles (except $T_+$) are { odd},
all { SM} particles are { even}. There are therefore
no contributions by { T odd} particles at the { tree level},
but the cancellation of divergences still works since it is a loop effect.
This allows lowering the scale 
$ f$ to $\sim 1 \,{\rm TeV}$ (or even lower).

The LHT model contains 
three doublets of { ``mirror quarks''} (T odd, heavy),
three doublets of { ``mirror leptons''} (T odd, heavy)
and a T odd $ T_-$ in addition to the T even $T_+$.
(Just like R parity in SUSY, T parity can also produce a
candidate for Dark Matter.)

The new parameters in the LHT model are
$ f$, the NP scale which also fixes $M_{W_H}$ etc.
The mixing between $t$ and $T$ is described by
  $ x_L$.
There are three  mirror quark masses: $ m_{H1}, m_{H2}$ and $ m_{H3}$ 
{ (the model is {\it Minimal Flavour Violating} (MFV) if these are degenerate)}
and a  mirror quark mixing matrix $ V_{Hd}$ 
containing three angles and three \cite{Blanke:2006xr} phases.
The up-type mirror fermion mixing matrix is given by
$V_{Hu}^\dagger V_{Hd}=V_{C\!K\!M}$.
(There are also 9 mirror lepton parameters, but these are not
of interest in the context of this study.)

\section{Flavour effects from LHT}

\begin{wrapfigure}{o}[2cm]{0pt}
\includegraphics[width=2.7cm]{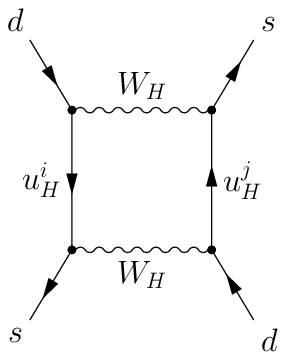}
\end{wrapfigure}
Although the { LHT} model does not introduce new operators
in addition to the SM ones, it is {\bf not} { MFV} because of
the mirror quark mixing. New particles contribute to 
Flavour Changing Neutral Current { (FCNC)} processes as
shown in the figure. A detailed discussion of Flavour Physics
in the LHT model is given in \cite{Blanke:2006eb}.

The LHT amplitudes can be written as (e.g.\ $K$ sector) \\
$\,\, \sum_{i=u,c,t} \,\,
{\lambda_i^K} F_i(m_i,m_{T^+},\dots)\,+\,
{ \xi_i^K }G_i(m_H^i,M_{W_H},\dots) $, where the 
first term is the T even contribution and the second term
is the T odd contribution. This way the Inami-Lim functions
become $ X_K = X_{\rm SM} +  X_{\rm even} + { \xi_i^{K}}/{ \lambda_t^K} 
  X_{\rm odd}$, with the CKM factors
$  { \lambda_t^{K}} = V_{ts}^*\,V_{td}$ and the mirror quark mixing
$ { \xi_i^{K}}=V^{*is}_{H\!d}V^{id}_{H\!d}$.
Because of the { CKM hierarchy}
$ 1/\lambda_t^K \gg 1/\lambda_t^{B_d} \gg 1/\lambda_t^{B_s}$,
we expect the { largest effects} in $ K$ physics,
 but suitable $ \xi_i^{j}$ can produce large effects also in $ B_d$, $ B_s$.

It has to be checked very carefully whether the LHT effects
do not violate existing experimental { FCNC constraints}.
We studied \cite{Blanke:2009am} the constraints on
$ \Delta M_K$ and $ \epsilon_K$ from the $K$ system,
the mass differences in the $B$ system
 $ \Delta M_{B_d}$ and $ \Delta M_{B_s}$, as well as the
CP asymmetry in $B_d$ decays $ S_{J\!/\!\psi K_{\!S}}$.
(Constraints from $ b\to s\gamma$ are not a problem,
the effects from LHT in this channel are very moderate.)

We generated { random points} in the LHT { parameter space},
checked these constraints and kept only points that { fulfill
all constraints}.
{The input parameters} were evenly distributed over their respective
{ $1\sigma$ ranges}.
\begin{wrapfigure}{r}[0cm]{0pt}
\includegraphics[width=7cm]{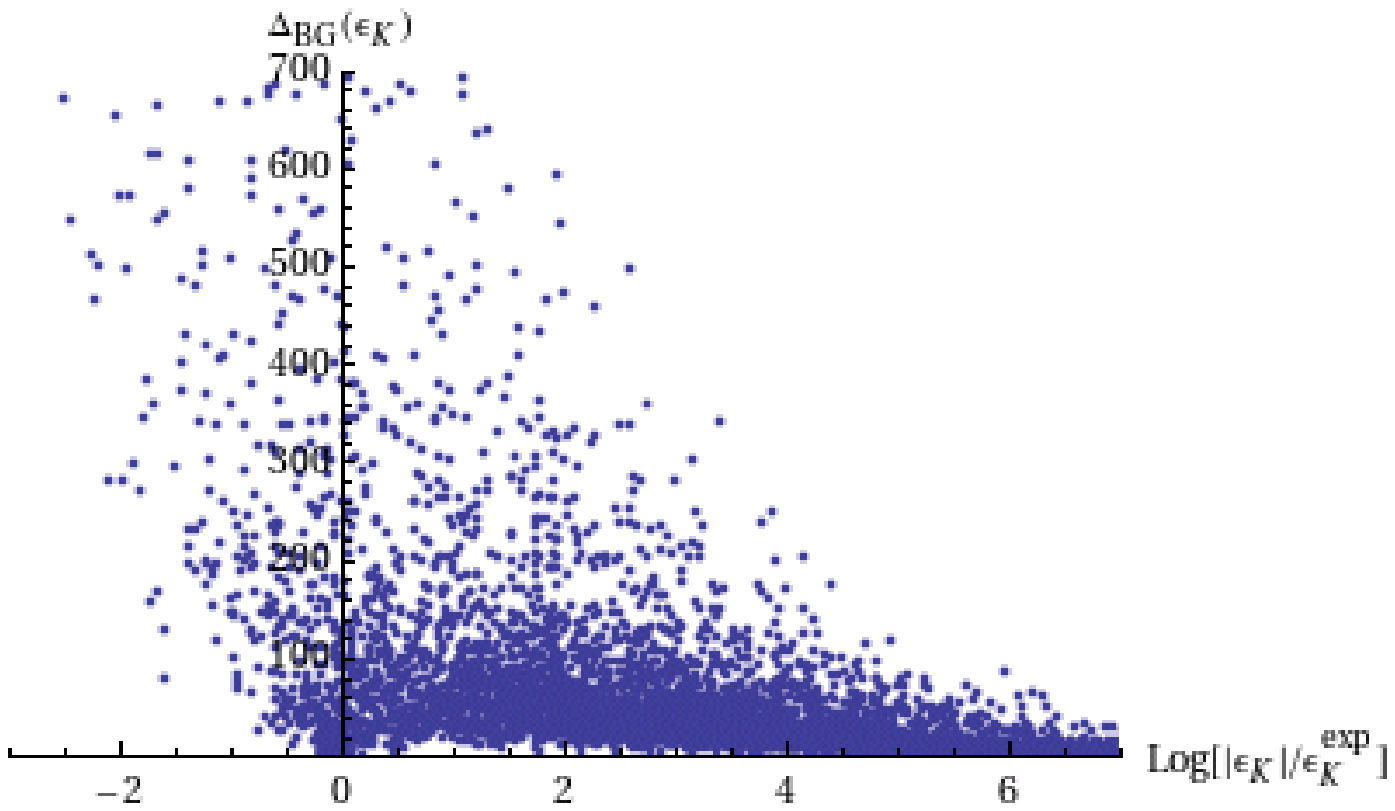}
\end{wrapfigure}
Although a lot of points in parameter space have to be tried
to find one that does not violate any of the experimental
constraints, fine tuning is not really a problem: Typically,
$\epsilon_K$ as generated by arbitrary model parameters is one
or two orders of magnitude too large, but there are also many
points that generate correct $\epsilon_K$ without large
fine tuning $\Delta_{\rm BG}(O)= {\rm max}_j  
\left| \frac{p_j}{O}\frac{\partial O}  {\partial p_j}\right| $
\cite{Barbieri:1987fn}. 
{ Some of the most spectacular points need { no fine tuning} at all.}

\section{ General results from LHT flavour study}

\begin{wrapfigure}{l}[0cm]{0pt}
\includegraphics[width=7cm]{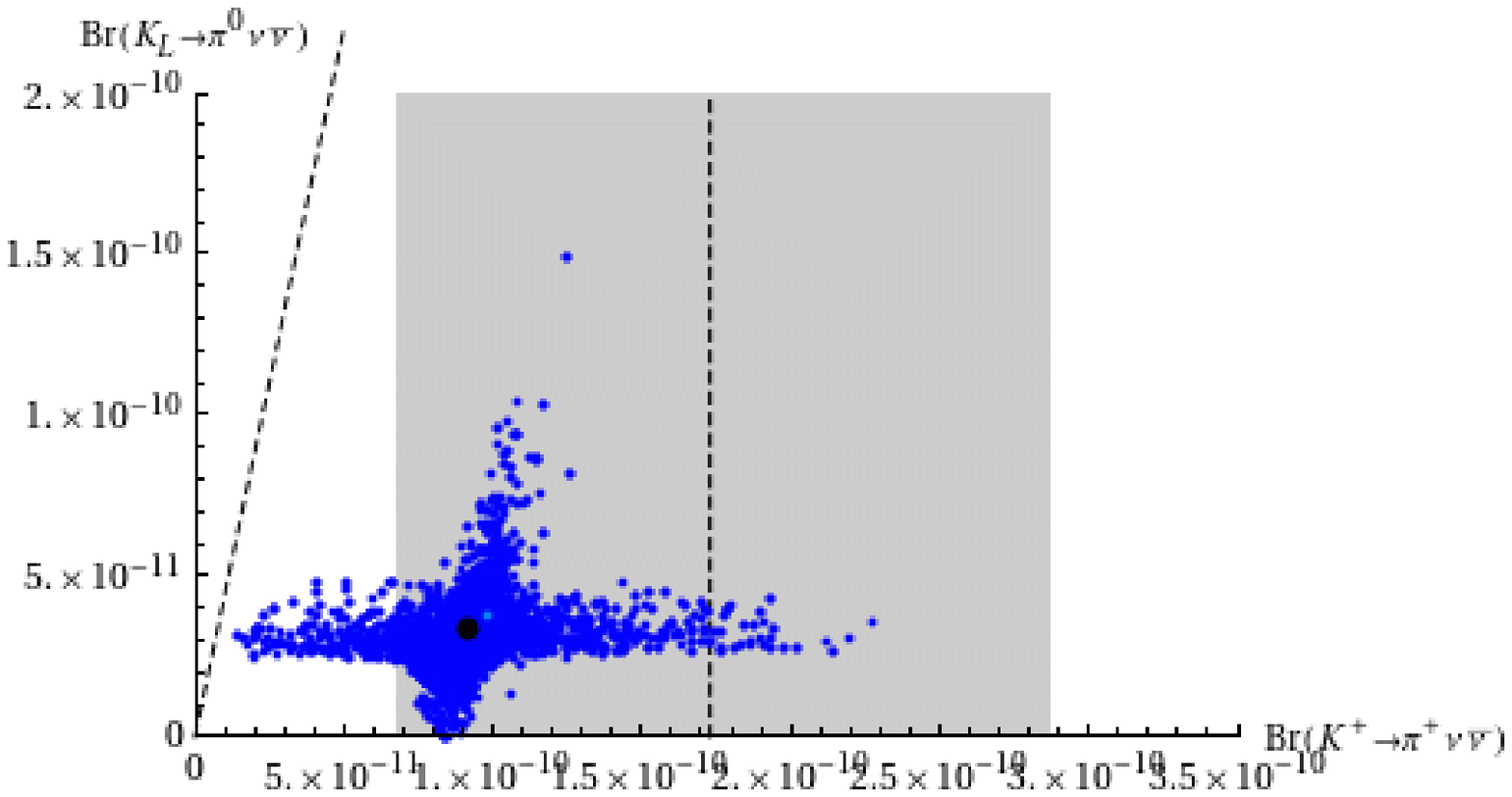}
\end{wrapfigure}
The decays  $ K^+\rightarrow\pi^+\nu\bar\nu$
and especially $ K_L\rightarrow\pi^0\nu\bar\nu$ are excellent
probes of new physics because they can be calculated very cleanly.
In the LHT model,
{ $K_L\rightarrow\pi^0\nu\bar\nu$ can be enhanced significantly over
the SM value (black dot) up to a factor of 3-5,
and also $ K^+\rightarrow\pi^+\nu\bar\nu$
can easily be enhanced to the central value (dashed line) of the
current experimental range.
Most data points lie on { two axes}: One of
constant $K_L\rightarrow\pi^0\nu\bar\nu$ and one
parallel to the Grossmann-Nir bound, this is due to the specific
{ operator structure} of the LHT model and distinguishes the
experimental signature from other models.


\mbox{}\hspace*{-.6cm}
\includegraphics[width=6cm]{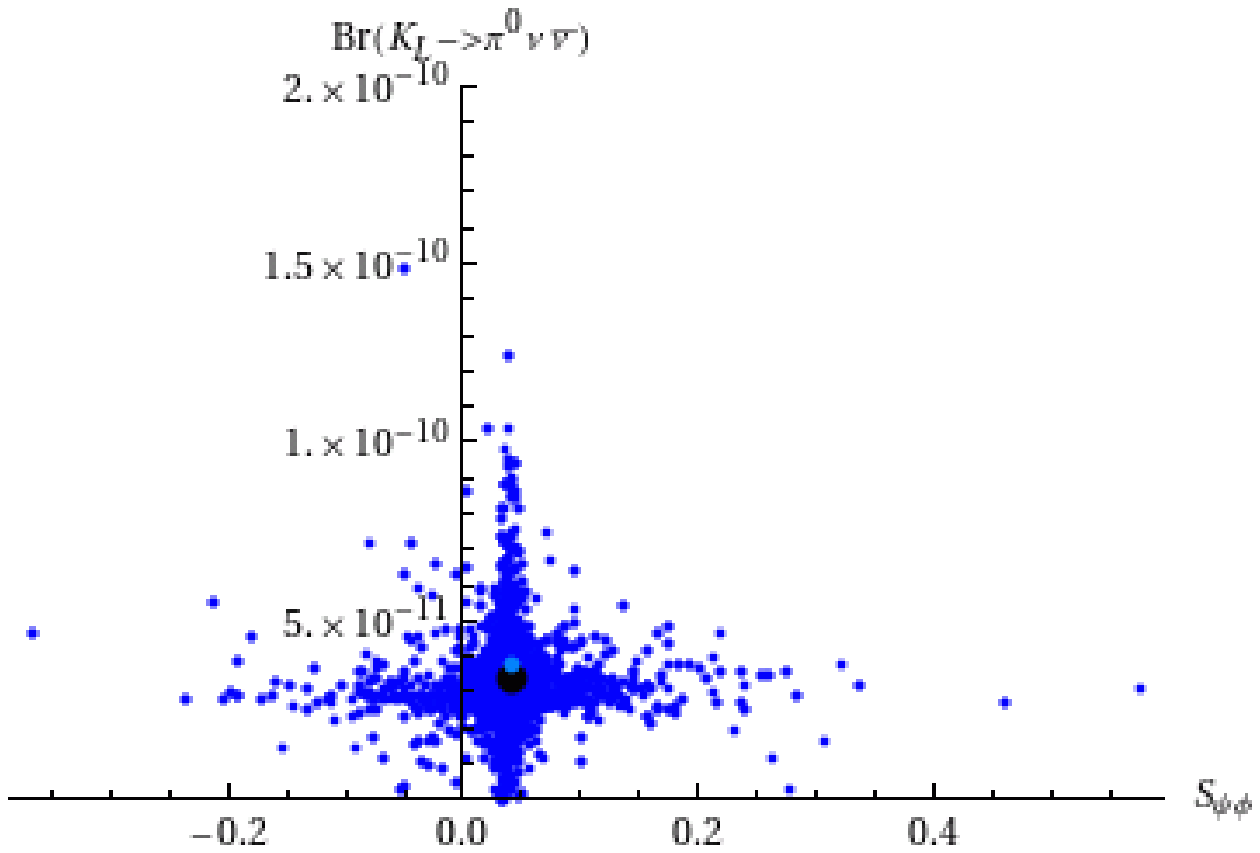}
\includegraphics[width=6.2cm]{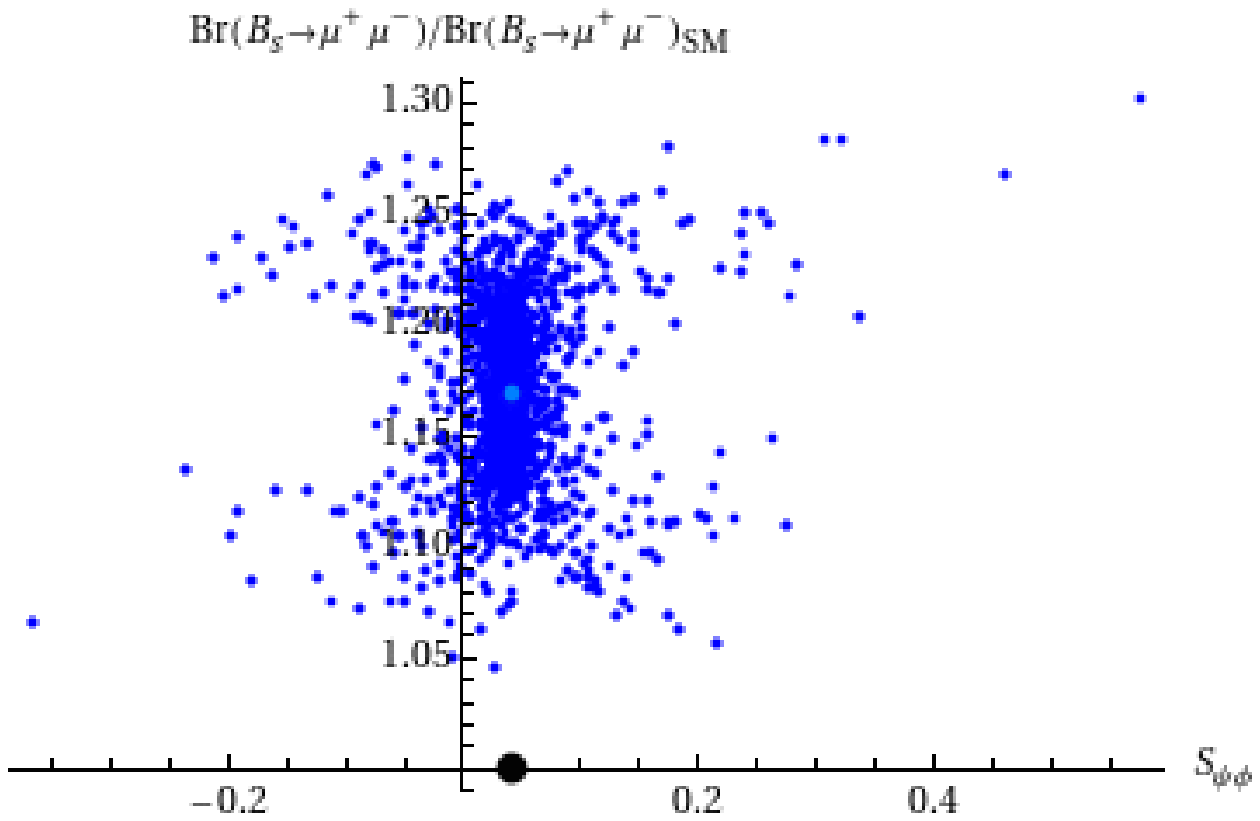}\\
The CP-asymmetry $ S_{\psi\phi}$ of the decay $B_s\to \psi\phi$
is much smaller in the SM than 
$ S_{J\!/\!\psi K_{\!S}}$ because the corresponding CKM angle
$\beta_s$ is only about $-1\deg$. In the LHT model, large
effects between -0.3 and +0.4 are observed, but 
{simultaneous large effects in $ K_L\rightarrow\pi^0\nu\bar\nu$ and
 $ S_{\psi\phi}$, though possible, seem unlikely.}
This is very different from the situation between
$ Br(B_s\to\mu^+\mu^-)$ and $ S_{\psi\phi}$, here simultaneous
significant effects are rather likely because both observables
profit from a modified $b\to s$ penguin. The enhancement of
$ Br(B_s\to\mu^+\mu^-)$ of up to 30\% over the SM result is,
however, rather moderate compared to e.g.\ SUSY.

\begin{wrapfigure}{r}[0cm]{0pt}
\includegraphics[width=7cm]{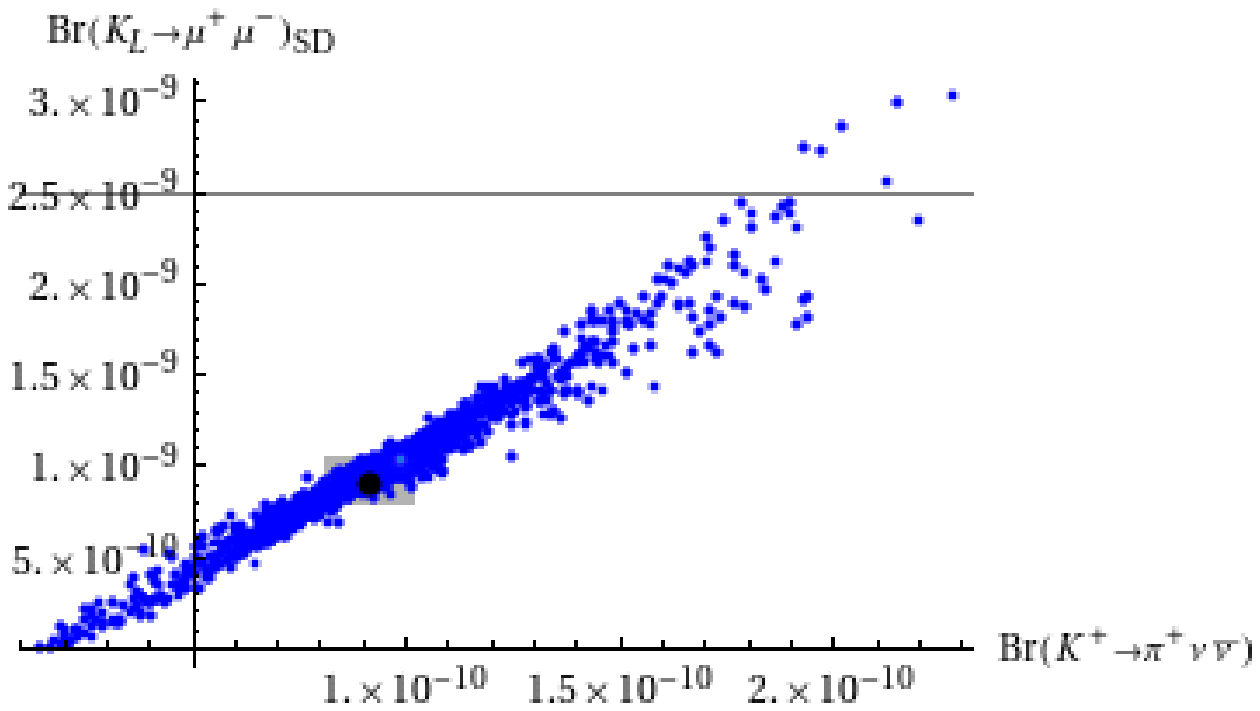}
\end{wrapfigure}
Another { interesting signature} of the LHT model is
the correlation between the Br's of
$ K_L\to\mu^+\mu^-_{\rm \,\,\,SD}$ and $ K^+\rightarrow\pi^+\nu\bar\nu$,
which is very different from e.g.\ the RS model with custodial
protection (c.f.\ contribution by B\"orn Duling in this volume).
Correlations like these might prove instrumental in distinguishing
different models of NP in the experiment.


\boldmath
\section{ $D\bar D$ Oscillations (in the LHT model) }
\unboldmath
\begin{wrapfigure}{o}[2cm]{0pt}
\includegraphics[width=3.5cm,bb=106 312 506 475]{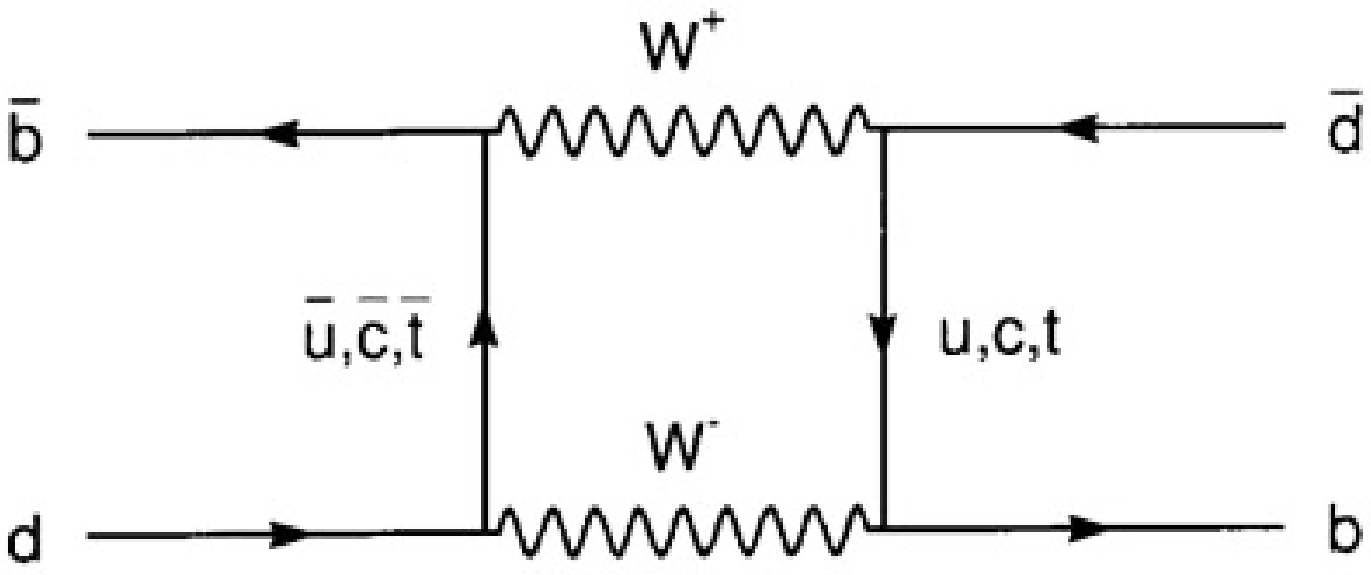}
\end{wrapfigure}
(This section is based on \cite{Blanke:2007ee,Bigi:2009df}.)
$D\bar D$ is more complicated than $K\bar K$ and $B\bar B$ mixing: 
$ K\bar K$ and $ B\bar B$ mixing is dominated by { short-distance
 physics, i.e.\ charm/top loops (c.g. figure).
$ D\bar D$ has almost { no short-distance} contribution:
 The corresponding CKM factors are small and the down-type 
quarks in the loops too light. Therefore the SM contribution
to $D\bar D$ mixing is { long-distance} and therefore
{ difficult to estimate}. In our analysis,
{ we vary the SM contribution} in a reasonable
 range and use theoretical estimates only to bound the values.

The $D$ mass eigenstates are $
| D_{1/2} \rangle = 1/\sqrt{|p|^2+|q|^2} 
   \left( p | D^0 \rangle \pm
 q | \bar D^0 \rangle \right)
$,
the { observables} are the normalised { mass} and { width differences}, 
$ x_D \equiv {\Delta M_D}/{\overline \Gamma} \, , 
 y_D \equiv {\Delta \Gamma_D}/{2\overline \Gamma} \, ,
$ as well as 
${q}/{p} \equiv \sqrt{({M_{12}^D}^* - \frac{i}{2}{{\Gamma_{12}^D}^*})/
  ({M_{12}^D - \frac{i}{2}\Gamma_{12}^D})}
$. Obviously CP is violated when $\left|q/p\right|\ne 1$.

{ Rather recently,{ $D\bar D$ oscillations have been
observed} \cite{Aubert:2007wf}, a measurement received with
great interest by the commmunity:
$
x_D = 0.0100^{+0.0024}_{-0.0026} \, , y_D = 0.0076^{+0.0017}_{-0.0018}
\,, 
\left|{q}/{p}\right| = 0.86 ^{+0.17}_{-0.15} \label{eq:qpexp}
$.
Although this establishes oscillation,
{ $CP$ violation has not (yet) been observed, 
{ $|q/p|$ is consistent with 1.}}
In the { SM}, no significant $CP$ violation is expected.

\begin{wrapfigure}{l}[0cm]{0pt}
\includegraphics[width=6cm]{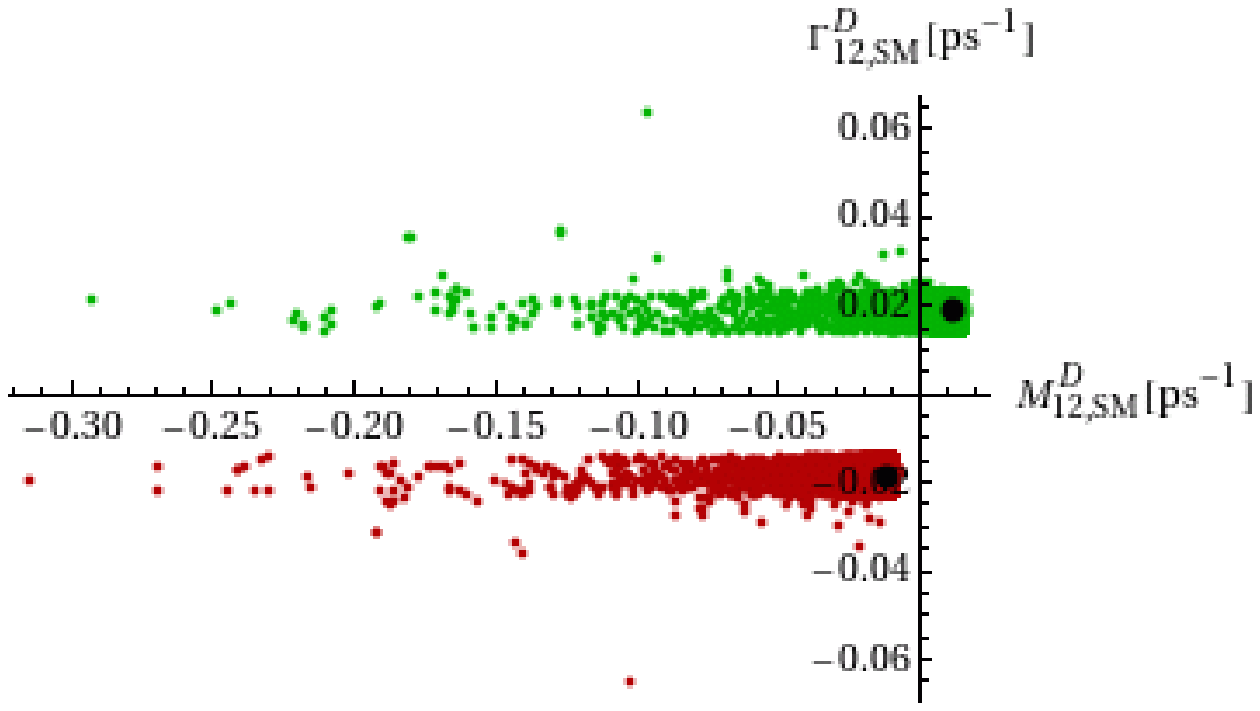}
\end{wrapfigure}
To establish whether the LHT model can produce a significant
$CP$ violation in the $D$ system,
we { determine} $ (M_{12}^D)_{\rm SM}$ and 
$ (\Gamma_{12}^D)_{\rm SM}$ so that { together with the LHT}
contribution, $ x_D$ and $ y_D$ coincide with experiment.
This approach is reasonable, because even the expected relative
sign of $ (M_{12}^D)_{\rm SM}$ and $ (\Gamma_{12}^D)_{\rm SM}$
\cite{Bigi:2000wn} does not match the values necessary to
reproduce the measured values of $x_D$ and $y_D$ with the
SM contributions, i.e.\ very little is known about these
quantities from the theoretical side. We obtain two solutions
for each LHT parameter point as shown in the figure.

Essentially all LHT parameter points are { consistent} with
{ expectations} for the magnitude of { SM contributions}.
In some cases,
$ (M_{12}^D)_{\rm SM}$ / $ (\Gamma_{12}^D)_{\rm SM}$ can 
be rather large, but these
are not our most spectacular/interesting data points.

\begin{wrapfigure}{r}[0cm]{0pt}
\includegraphics[width=5.5cm]{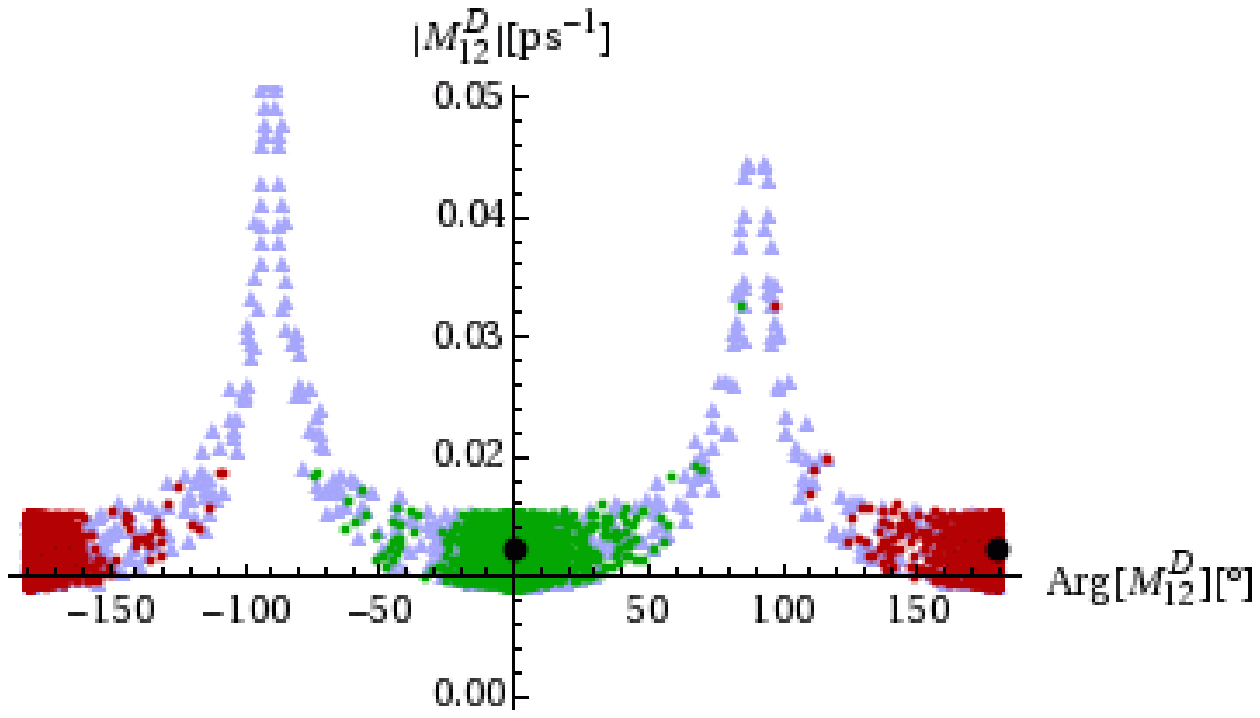}
\end{wrapfigure}
Obviously, requiring $x_D$ and $y_D$ to coincide with
experiment restricts the allowed points to a rather narrow
region in the ${\rm Abs}/{\rm Arg} M_{12}^D$ plane. Since
$V_{Hu}^\dagger V_{Hd}=V_{C\!K\!M}$ and the CKM-matrix is
rather close to the unity matrix, the experimental constraints
on { $\epsilon_K$ } exclude points with large ${\rm Arg} M_{12}^D$
(light blue/grey triangles).

\begin{wrapfigure}{l}[0cm]{0pt}
\includegraphics[width=6cm]{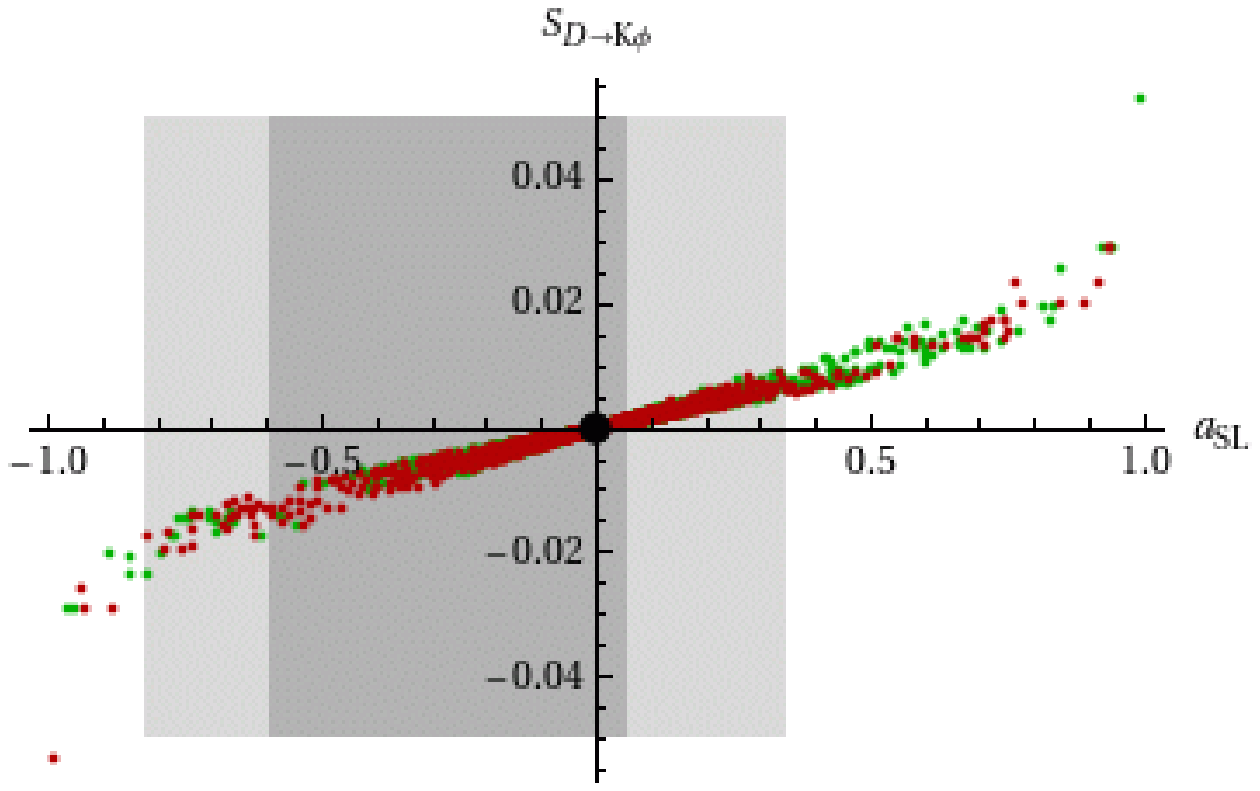}
\end{wrapfigure}
Even without these points, i.e.\ observing all
experimental constraints, very large (for the $D$ system)
CP asymmetries of several percent are possible. The LHT
model could even generate asymmetries of $\pm 5\%$ for
$D\to K\phi$, but this would correspond to semileptonic
asymmetries $a_{\rm SL}^D$ close to unity. Such large
values of $a_{\rm SL}^D$ are already excluded by the
measurements of  $ \left| q/p\right|_{\rm exp} = 0.86 ^{+0.17}_{-0.15}$
because $a_{\rm SL}^D = {(|q|^4\!-\! |p|^4)}/{(|q|^4 \!+\! |p|^4)}$. 
We can therefore conclude
that the LHT model can easily saturate the CP violation
in the $D$ system that is still allowed by current
measurements.

\begin{wrapfigure}{r}[0cm]{0pt}
\includegraphics[width=5.5cm]{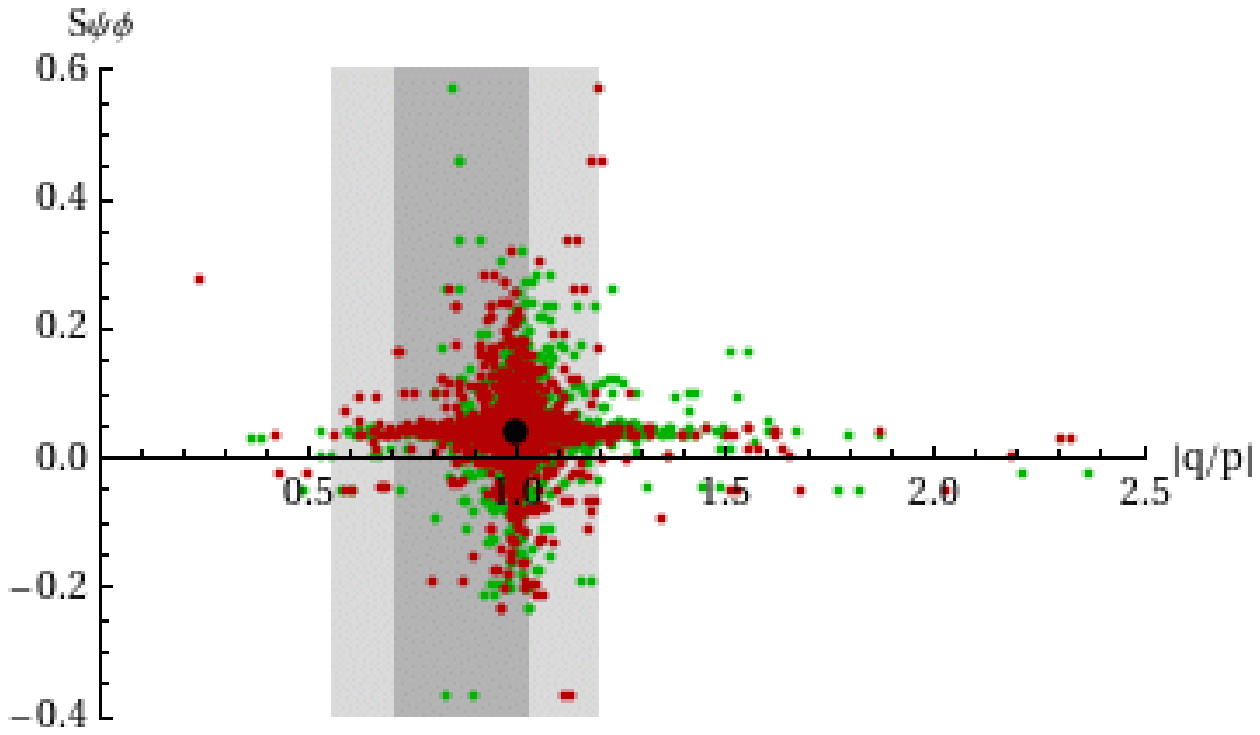}
\end{wrapfigure}
Let us last look at the correlation between the $D$ system
and the $B_s$ system: We find that simulataneous large NP
effects in both systems are possible, but unlikely, just
as we found that simultanous large effects in the $K$ and the $B$
system are unlikely in the LHT model. Again, it is easier to
produce large NP effects that do not violate existing experimental
constraints in one sector than in two.

\section{ Conclusions}

The  LHT model is an interesting, economical
 alternative to SUSY etc.\ in solving the  Little Hierarchy problem.
There are rather  few parameters, the model passes the EW precision  
tests and (surprisingly, because this is not what the model
was created for) there  are interesting, sometimes spectacular
 effects on  Flavour observables.
For example, large $ CP$ violation in { $D\bar D$ oscillations} 
is possible.
We hope that in the near future, experimental results will
show us whether nature has chosen anything like the LHT model
for physics at the TeV scale.

\end{document}